# Brushing Feature Values in Immersive Graph Visualization Environment

Hinako Sassa[1)]   Maxime Cordeil[2)]   Mitsuo Yoshida[3)]   Takayuki Itoh[1)]

1) Ochanomizu University   2) Monash University   3) Toyohashi University of Technology

**ABSTRACT**

There are a variety of graphs where multidimensional feature values are assigned to the nodes. Visualization of such datasets is not an easy task since they are complex and often huge. Immersive Analytics is a powerful approach to support the interactive exploration of such large and complex data. Many recent studies on graph visualization have applied immersive analytics frameworks; however, there have been few studies on immersive analytics for visualization of multidimensional attributes associated with the input graphs. This paper presents a new immersive analytics system that supports the interactive exploration of multidimensional feature values assigned to the nodes of input graphs. The presented system displays "label-axes" corresponding to the dimensions of feature values, and "label-edges" that connect label-axes and corresponding to the nodes. The system supports brushing operations which controls the display of edges that connect a label-axis and nodes of the graph. This paper introduces visualization examples with a graph dataset of Twitter users and reviews by experts on graph data analysis.

## 1 INTRODUCTION

Many graph datasets in our daily life have feature values assigned to their nodes. For example, scores of exams can be described as feature vectors of friendship networks of students. Or, the numbers of listened tunes for each music genre can be described as feature vectors of listeners network of on-line music services. Similarly, structured datasets can be produced in various academic and industrial fields including bioinformatics and traffic analysis. The recent evolution of machine learning techniques including graph neural network [24] made this type of datasets useful and valuable. Also, many recent graph visualization studies [11,18] addressed the representation of such datasets.

Suppose that we have a dataset of listeners network, where each listener has numbers of listened tunes for each music genre as a feature vector. Nodes correspond to listeners, edges are friendships between pairs of listeners, and dimensions of feature vectors correspond to genres. We may want to know which genres of tunes are most frequently listened by which portions of the network. Or, we may want to know how many listeners accessed each genre of tunes. In other words, the following are typical requirements for visualization of this type of datasets:

**[R1:]** Distribution of the dominant dimension of each node in the network structure.

**[R2:]** Distribution of feature values of a particular dimension and its relationship with the network structure.

Developments of visualization techniques for the above requirements is not an easy problem since the datasets are complex and often huge. Immersive analytics [17] is a powerful approach to interactively explore and analyze such complex data. There have been several recent graph visualization studies that applied immersive technologies such as virtual reality (VR) devices; however, these studies did not focus on the visualization of feature values of the graphs.

This paper presents a new immersive analytics framework that supports interactive explorations of multidimensional attributes associated with the input graphs. This system is developed on the top of ImAxis [5], an immersive analytics toolkit for multidimensional datasets. The presented system displays "label-axes" corresponding to the dimensions of feature vectors as original ImAxis displays independent axes to represent the dimensions of input datasets. Also, the system displays "label-edges" that connect a label-axis and nodes of the graph to represent how the distribution of a particular dimension of the feature values relates to the graph structure. The main contribution of this study is supporting "brushing operation" which controls the display of label-edges in an immersive space.

This paper introduces a case study with a network dataset of Twitter retweet network of tweets of political parties and expert reviews of network analysis while using the presented system.

## 2 RELATED WORK

Many real-world graph data have various attributes of nodes. One of the typical attributes is a set of labels. Set visualization techniques [1] have been well-applied to represent graphs with such labels. Euler diagram or region-based overlay representations are especially well-applied to graphs [3,9,13]. Icons and glyphs are also effective approaches for the representation of labels [12]. Hybrid techniques [6] have been also discussed. Feature vectors which have multidimensional real values are another typical attributes assigned to the nodes of the graphs [19,20]. The similarity of feature vectors can be reflected in the layout of node-link diagrams [11,18]; however, it is generally difficult to finely represent the multidimensional values in a node-link diagram drawn in a limited display space. It is desirable to implement linked views or other interactive mechanisms to finely represent the multidimensional feature values as well we the node-link diagram.

Immersive analytics [17] is a new framework for interactive and intuitive data analysis. Several software toolkits for immersive analytics have been recently released. ImAxis [5] is a typical one mainly for interactive analysis of multidimensional data. Users can place the axes corresponding to the dimensions of the data anywhere in the virtual space while using ImAxis with virtual reality devices. The tool provides various multidimensional data visualization techniques including PCP (Parallel Coordinate Plots) and SPLOM (Scatterplot matrices) with arbitrary sets of dimensions according to the manipulations of the axes. There have been several newer studies on immersive multidimensional data visualization [8,21], development of more robust immersive analytics systems [4], and evaluation of immersive analytics processes [2]. The presented immersive graph visualization has been developed on the top of ImAxis so that users can interactively and immersively explore multidimensional values associated with the nodes of the graph.

There have been several studies on immersive graph analytics. Kwon et al. [15] discussed graph layout techniques on sphere and routing, bundling and rendering of edges for immersive graph visualization. Huang et al. [10] discussed the effectiveness of gestures for immersive graph visualization. Several evaluations by user studies on immersive graph visualization [7,14,16] have been also presented. However, there have been few studies that apply immersive analytics frameworks for interactive exploration of multidimensional attributes associated with the input graphs.

## 3 IMMERSIVE GRAPH VISUALIZATION WITH BRUSHING FEATURE VALUES

This section presents the technical details of our immersive graph visualization implemented on ImAxis [5] and its main feature on brushing feature values assigned to nodes of the input graph.

### 3.1 Data structure

This system supposes that all nodes of the input graph have feature vectors. Let us formalize the input graph $G=(V,E)$ where $V$ and $E$ are nodes and edges respectively. The set of nodes $V$ are defined as $V=\{v_1, v_2, ..., v_n\}$, where $n$ is the number of nodes. The $i$-th node $v_i$ has the following variables:

$v_i=\{x_i, y_i, z_i, a_i\}$, where $x_i$, $y_i$ and $z_i$ denote the position in a 3D space, and $a_i = \{a_{i1}, a_{i2}, ..., a_{im}\}$ is the feature vector, where $m$ is the number of dimensions.

The system displays the graph and axes corresponding to dimensions of the feature vectors. In addition to displaying nodes and edges, the system draws the segments that connect nodes and the axes of dimensions. This paper calls the axes as "label-axes" and the segments as "label-edges". The number of displayed label-edges is controlled by the brushing operations on the label-axes.

### 3.2 Layout and drawing of the input graph

The system can apply for precomputed node positions or compute the layout by our own implementation. Our implementation firstly calculates the following distances defined by Itoh et al. [11] between arbitrary pairs of nodes, where $a_i$ and $a_j$ are feature vectors of the $i$-th and $j$-th nodes:

$$d = 1.0 - \frac{a_i \cdot a_j}{|a_i||a_j|}$$

Then, the implementation applies dimension reduction schemes to the distance matrix to calculate the positions of the nodes. We currently apply multidimensional scaling (MDS) for dimension reduction. As a result, nodes of those feature vectors are similar are placed closer.

The system draws the graph in a 3D space. Colors of nodes are assigned according to the dimension that has the largest values in their feature vectors as shown in Figure 1, where edges connecting a pair of nodes are drawn with the gradation of two colors of the nodes. The sizes of nodes and thickness of edges are fixed. Users can recognize the distribution of dominant dimension and subareas where particular dimensions have larger values (**R1**).

### 3.3 Display of label axes

Figure 1 also illustrates how the label-axes are generated from the input graph. A label-axis for the $k$-th dimension is generated from a set of feature values of the $k$-th dimension $A_k$ = $\{a_{1k}, a_{2k}, ... a_{mk}\}$. The maximum value of $A_k$ is assigned at the upper end of the label-axis, and the minimum value is at the lower end. A histogram that represents the distribution of $A_k$ is displayed with the label-axis. Users can grab, release, and throw the label-axes using the controller devices. The label-axes has a filtering function to narrow down the ranges of $A_k$. The ranges are represented by black and white rhombus objects. Users can freely move these objects to specify the ranges.

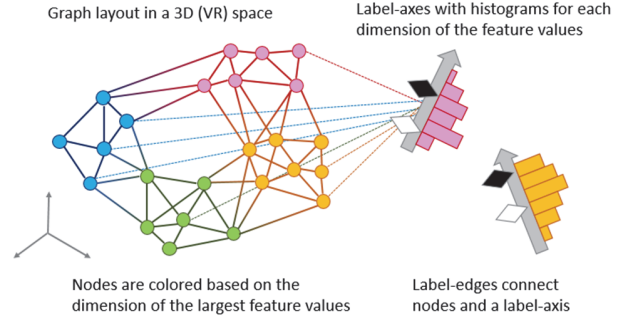

Figure 1: Illustration of the brushing operation.

### 3.4 Brushing

The system displays label-edges between nodes and a label-axis when a user moves the label-axis sufficiently close to the graph. We call this operation "brushing" because a set of label-edges may look like a brush. The label-edges are drawn as dotted lines because it is easier to visually distinguish with the edges of the graph. This interaction assists users to observe the relations between the distribution of feature values of particular dimensions and graph structure (**R2**).

Users can briefly understand the real values of the particular dimension corresponding to the user-operated label-axis by looking at the label-edges. Figure 1 illustrates how label-edges connect the nodes and the user-operated label-axis. Here, many label-edges connected to the upper side of the label-axis are drawn in blue or green. It suggests that the dimension corresponding to the current label-axis is correlated with the other two dimensions corresponding to blue and green. Users can interactively explore the correlation between the dimension corresponding to the user-operated label-axis and other dimensions.

### 3.5 Running Environment

We developed this system using the game engine Unity and the VR device HTC Vive. This environment enables flexible viewpoint control and interactive label-axis operations simultaneously since HTC Vive equips a head-mount display and two controllers.

## 4 EXAMPLES

This section introduces examples with a network dataset of Twitter users [22,23]. The dataset records Twitter users who retweet the tweets by official accounts of Japanese political parties and forms a network by traversing the flow of retweets and connecting the Twitter users. Then, we extracted a subgraph consisting of 1000 Twitter users and treated the numbers of retweeted tweets for each of the parties as feature vectors of Twitter users. Figure 2 shows the eight official Twitter accounts of political parties contained in this dataset and the colors assigned to these parties. Nodes corresponding to the Twitter accounts are colored according to the political party that the users most frequently retweeted. As a result, nodes in this dataset have eight-dimensional feature vectors, and therefore, eight label-axes are generated in the visualization results.

- 🟥 @jimin_koho
- 🟧 @CPD2017
- 🟨 @nipponnokokoro
- 🟩 @osaka_ishin
- 🟦 @jcp_cc
- 🟦 @komei_koho
- 🟪 @hr_party_TW
- 🟪 @kibounotou

Figure 2: Political parties and these colors.

Figure 3 demonstrates typical efforts of the brushing operations for the visualization of the correlation between the labels and the graph structure. Figure 3(left) shows the label-edges with @nipponkokoro (The Party for Japanese Kokoro), while Figure 3(right) shows the label-edges with @jimin_koho (Liberal Democratic Party). Here, many label-edges connecting the upper side of the label-axis in Figure 3(left) are colored in red, while many other label-edges connecting the lower side are colored in yellow. As shown in Figure 2, @nipponkokoro is colored in red, while @jimin_koho is colored in yellow. This result suggests that users that mainly retweet the tweets by @jimin_koho also often retweet the tweets by @nipponkokoro. Similarly, users that mainly retweet the tweets by @nipponkokoro also often retweet the tweets by @jimin_koho. Actually, these two parties argued similar politics and finally merged.

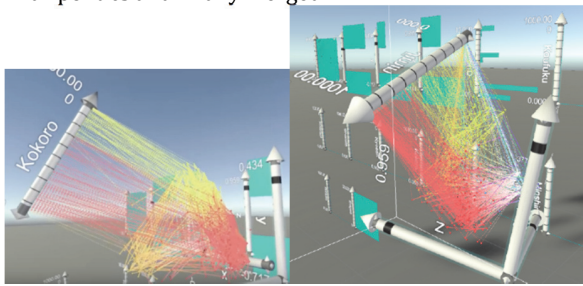

Figure 3: Brushing operations. (left) Specifying @nipponkokoro. (right) Specifying @jimin_koho.

Our system enables complex brushing operations with multiple label-axes. Figure 4 shows an example of brushing operations with two label-axes corresponding to @jimin_koho and @nipponkokoro. A particular set of Twitter users who actively retweet the tweets of both @jimin_koho and @nipponkokoro can be discovered by this operation.

Figure 5 shows the brushing operation with the label-axis of @komei_koho (Komeito). Many label-edges colored in blue, red, and yellow appeared during this operation. Here, blue corresponds to @komei_koho, red corresponds to @jimin_koho, and yellow corresponds to @nipponkokoro as shown in Figure 2. This result suggests that Twitter users who retweet the tweets of @komei_koho are also active to retweet the tweets of @jimin_koho and @nipponkokoro. We estimate one of the reasons is that both @komei_koho and @jimin_koho were ruling parties at that time.

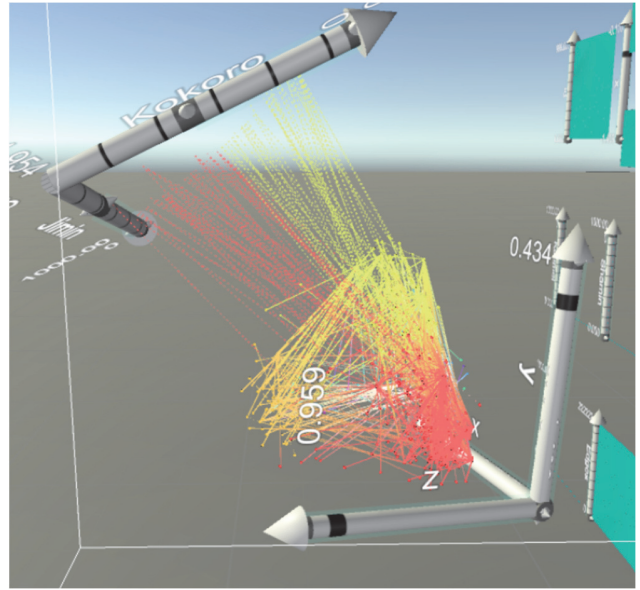

Figure 4: Brushing operation with two label-axes.

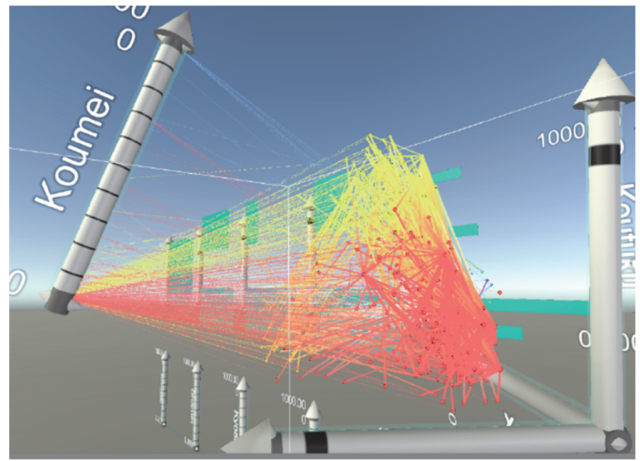

Figure 5: Brushing operation with the label-axis of @komei_koho.

Figure 6 shows the brushing operation with the label-axis of @jcp_cc (Japanese Communist Party). Many label-edges colored in sky-blue, purple, and orange appeared during this operation. A small number of red label-edges corresponding to @jimin_koho also appeared. Here, sky-blue corresponds to @jcp_cc, purple corresponds to @kibounotou (Party of Hope), and orange corresponds to @CDP2017 (The Constitutional Democratic Party of Japan) as shown in Figure 2. Also, Figure 6 shows that many sky-blue label-edges connect to the upper part of the label-axis, many purple label-edges are in the middle part, while orange and red label-edges connect to the lower part. This result suggests @kibounotou is the most related to @jcp_cc. @CDP2017 is secondly, and @jimin_koho is thirdly correlated to @jcp_cc. We found that @jcp_cc and @CDP2017 collaborated as an opposition party group at that time and often had tweets of objections to @jimin_koho. Also, we found that @jcp_cc negotiated with @kibounotou to merge each other at that time.

Figure 7 shows the brushing operation with the label-axis of @CDP2017. Many sky-blue and red label-edges corresponding to @jcp_cc and @jimin_koho appear as well as orange label-edges in this visualization. Both @jcp_cc and @CDP2017 were opposition parties while @jimin_koho was the largest ruling party. We estimate that many Twitter users who are supporters of @CDP2017 want to compare the argues of @CDP2017 with those of @jcp_cc as a major opposition party and @jimin_koho as the largest ruling party.

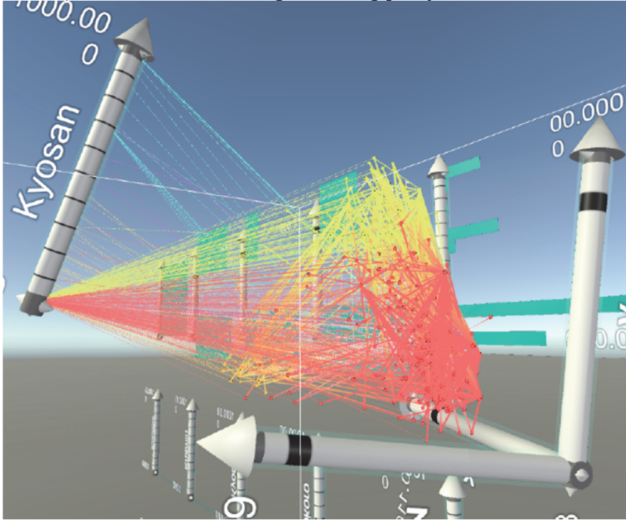

Figure 6: Brushing operation with the label-axis of @jcp_cc.

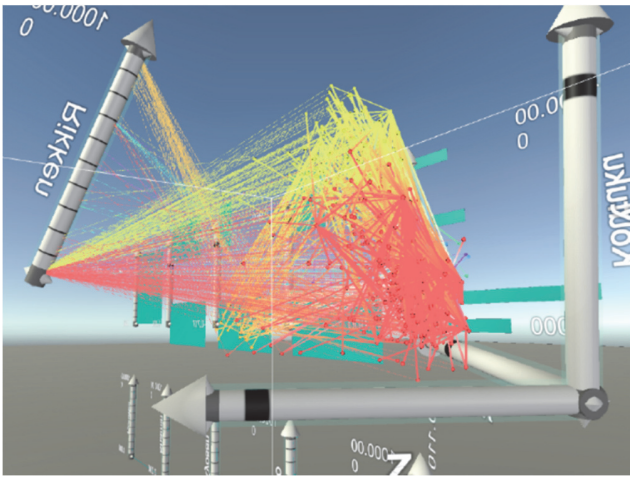

Figure 7: Brushing operation with the label-axis of @CDP2017.

The above visualization results show that many Twitter users retweet the tweets of multiple parties because label-axes are connected to multiple colors of label-edges. Also, a large number of users supporting various parties retweets the tweets of @jimin_koho because red label-edges appeared from all the label-edges. We found many Twitter users retweet the tweets of collaborating parties such as @jcp_cc and @CDP2017 while many others retweet the tweets of opposed parties such as @CDP2017 and @jimin_koho.

## 5 EXPERT REVIEW

We invited two experts on network analysis to our experiment. We observed how they observe the network dataset using our system and discussed the effectiveness of our system. We used the Twitter network dataset introduced in the previous section for this experiment.

### 5.1 Procedure

We explained the participants the procedure of this experiment, functionality of the system, and the detail of the Twitter network dataset. Then, we asked the participants to practice the operations of the system for five minutes while we advised the operations. Also, we asked them to estimate which parties are eager to promote themselves via Twitter. Participants observed the dataset to discover the interesting behavior of the parties using this system for one hour. We asked them to explain to us why they wanted to have the current operations one-by-one. Finally, participants answered the questionnaire including the following questions.
1. Past experiences of VR systems
2. Difficulty of operations with HTC Vive
3. Comprehensibility of visualization with brushing operations
4. Discomfort while using this system
5. Free feedback

### 5.2 Results

#### 5.2.1 Analysis of the political parties

One of the participants estimated that @osaka_ishin (Japan Innovation Party) and @CDP2017 are eager to promote themselves via Twitter. She found that red and yellow label-edges corresponding to @jimin_koho and @nipponkokoro are connected to most of the label-axes. This means that many Twitter users who mainly retweet the tweets of @jimin_koho and @nipponkokoro also retweet the tweets of almost all parties; in other words, they are not focused on supporting @jimin_koho and @nipponkokoro. Therefore, she did not think @jimin_koho and @nipponkokoro are successful to promote themselves via their own Twitter accounts. Instead, she focused on some colors of label-edges intensively connected to particular label-axes. Green and orange label-edges corresponding to @osaka_ishin and @CDP2017 were typical ones; therefore, she estimated these two parties are successful to promote their Twitter accounts.

On the other hand, another participant estimated @jimin_koho and @nipponkokoro were successful to promote themselves via their own Twitter account. She thought these two parties are most famous because the largest number of label-edges corresponding to these parties are connected to most of the label-axes.

#### 5.2.2 Answers for questionnaire

Both two participants answered that this system was easy to use even though they did not have any experience with data analysis with VR systems. Also, they answered feature vector representation with the brushing operation was easy to understand. They did not feel any VR sickness while the experiments.

The following are problems the participants mentioned. Fixed sensors sometimes prevented the operations of controllers. It was difficult to remember the assignment of colors to political parties. Histograms on the label-axes were too small. The frame rate was not sometimes sufficient.

The participants also suggested developing additional functions such as detail-on-demand display of node information and scaling/fisheye of the graph.

### 5.3 Discussion

We found that two participants had a different understanding of the promotion of Twitter accounts. In other words, we found our system enables flexible interactions for users who have a different understanding of the same dataset.

We were afraid that the brushing operation was not a common one and therefore it might be difficult to master for novice users. However, the participants mentioned it was easy to master. It suggests that the brushing operation developed in our system and intuitive and easy to use for a wide range of users.

## 6 CONCLUSION AND FUTURE WORK

This paper presented a new immersive analytics framework for interactive exploration of multidimensional attributes associated with the input graphs. The presented system displays label-axes corresponding to the dimensions of feature vectors as original ImAxis displays independent a. The system supports the brushing operations which controls the display of label-edges that connect a label-axis and nodes of the graph. The paper introduced examples with the network datasets of Twitter users who retweet the tweets by official accounts of Japanese political parties. Also, we verified the usefulness of the system by the expert reviews with two researchers on graph analysis.

We would like to improve the scalability of the implementation. The visualization examples have heavy clutterings even the dataset has just 1000 nodes. We would like to apply conventional techniques including multi-layer node representation and edge bundling to the system.

This paper introduced our expert review that verifies how the system is useful for graph data analysis. Meanwhile, it is important to verify how the implemented immersive and interactive mechanisms are easy-to-use. We would like to conduct user experiments to verify the usability of the immersive and interactive mechanisms, and then improve the implementation based on the experimental results.

After these improvements, we would like to apply various datasets as well as the Twitter dataset.